%% file: sample-sigconf.tex
\documentclass[sigconf,natbib=true,camera-ready]{acmart}
\copyrightyear{2025}
\acmYear{2025}

\acmConference[SIGIR '25]{the 48th International ACM SIGIR Conference on Research and Development in Information Retrieval}{July 13-18, 2025}{Padua, Italy}

\AtBeginDocument{%
  }

\usepackage[most]{tcolorbox}
\usepackage{tabularx}
\usepackage{bbm}
\usepackage{hyperref} 
\usepackage{multirow}
\usepackage{makecell}
\usepackage{balance}
\usepackage{pifont}
\newcommand{\cmark}{\ding{51}}%
\newcommand{\xmark}{\ding{55}}%
\begin{document}

\title{U-Sticker: A Large-Scale Multi-Domain User Sticker Dataset for Retrieval and Personalization}

\author{Heng Er Metilda Chee}
\email{xxe23@mails.tsinghua.edu.cn}
\affiliation{%
\institution{DCST, Tsinghua University, Beijing, China. Quan Cheng Laboratory, Jinan, China.}
  \city{}
  \country{}
}

\author{Jiayin Wang}
\email{JiayinWangTHU@gmail.com}
\affiliation{%
  \institution{DCST, Tsinghua University}
  \city{Beijing}
  \country{China}
}



\settopmatter{printacmref=true}


\copyrightyear{2025}
\acmYear{2025}
\setcopyright{cc}
\setcctype{by}
\acmConference[SIGIR '25]{Proceedings of the 48th International ACM SIGIR
Conference on Research and Development in Information Retrieval}{July 13--18,
2025}{Padua, Italy}
\acmBooktitle{Proceedings of the 48th International ACM SIGIR Conference on
Research and Development in Information Retrieval (SIGIR '25), July 13--18,
2025, Padua, Italy}\acmDOI{10.1145/3726302.3730311}
\acmISBN{979-8-4007-1592-1/2025/07}

\author{Zhiqiang Guo}
\email{georgeguo.gzq.cn@gmail.com}
\affiliation{%
  \institution{DCST, Tsinghua University}
  \city{Beijing}
  \country{China}}

\author{Weizhi Ma}
\authornote{Corresponding authors. 
\\This work is supported by the Natural Science Foundation of China (Grant No. U21B2026, 62372260), Quan Cheng Laboratory (Grant No. QCLZD202301).
}
\email{mawz@tsinghua.edu.cn}
\affiliation{%
\institution{AIR, Tsinghua University}
 \city{Beijing}
 \country{China}}

\author{Qinglang Guo}
\email{gql1993@mail.ustc.edu.cn}
\affiliation{%
\institution{CETC Academy of Electronics and Info Tech Group Co.,Ltd.; China Academic of Electronics and Info Tech}
 \city{Beijing}
 \country{China}}
 
\author{Min Zhang}
\authornotemark[1]
\email{z-m@tsinghua.edu.cn}
\affiliation{%
  \institution{DCST, Tsinghua University, Beijing, China. Quan Cheng Laboratory, Jinan, China.}
  \city{}
  \country{}
  }

\makeatletter
\gdef\@copyrightpermission{
  \begin{minipage}{0.2\columnwidth}
   \href{https://creativecommons.org/licenses/by/4.0/}{\includegraphics[width=0.90\textwidth]{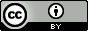}}
  \end{minipage}\hfill
  \begin{minipage}{0.8\columnwidth}
   \href{https://creativecommons.org/licenses/by/4.0/}{This work is licensed under a Creative Commons Attribution International 4.0 License.}
  \end{minipage}
  \vspace{5pt}
}
\makeatother



\renewcommand{\shortauthors}{Heng Er Metilda Chee et al.}


\begin{abstract}

Instant messaging with texts and stickers has become a widely adopted communication medium, enabling efficient expression of user semantics and emotions. With the increased use of stickers conveying information and feelings, sticker retrieval and recommendation has emerged as an important area of research. However, a major limitation in existing literature has been the lack of datasets capturing temporal and user-specific sticker interactions, which has hindered further progress in user modeling and sticker personalization.
To address this, we introduce \textbf{U}ser-\textbf{Sticker}, a dataset that includes temporal and user anonymous ID across conversations. It is the largest publicly available sticker dataset to date, containing {22K unique users}, {370K stickers}, and {8.3M messages}. The raw data was collected from a popular messaging platform from {67} conversations over 720 hours of crawling. All text and image data were carefully vetted for safety and privacy checks and modifications.
Spanning {10 domains}, the U-Sticker dataset captures rich {temporal}, {multilingual}, and {cross-domain behaviors} not previously available in other datasets. 
Extensive quantitative and qualitative experiments demonstrate U-Sticker's practical applications in user behavior modeling and personalized recommendation and highlight its potential to further research areas in personalized retrieval and conversational studies.
U-Sticker dataset is publicly available\footnote{https://huggingface.co/datasets/metchee/u-sticker}.



\end{abstract}





\maketitle

\input{Sections/S1-introduction}
\input{Sections/S2-related_work}

\input{Sections/S3-method}
\input{Sections/S4-experiments}

\clearpage

\bibliographystyle{ACM-Reference-Format}
\balance
\bibliography{sample-base}

\appendix
\input{Sections/S5-appendix}

\end{document}

%% file: Sections/S1-introduction.tex
\section{Introduction}
\label{sec:intro}
Instant messaging (IM) has become an essential mode of communication, allowing users to interact efficiently. Beyond text, stickers have emerged as a powerful medium of expression, enabling users to convey semantics and emotions comfortably and accurately~\cite{stickerclip,MOD}.
Unlike general image datasets, which emphasize object recognition, stickers carry rich semantic meaning, expressiveness, and emotional nuance, making their retrieval and recommendation fundamentally different from traditional image retrieval and recommendation tasks. 

However, despite stickers' widespread use, personalized sticker retrieval and recommendation remains an understudied area, primarily due to the lack of large-scale user sticker interaction dataset.
As shown in Table~\ref{tab:dataset_comparison}, while several sticker dataset exist, they often lack user information, making them unable to identify the same user across different conversations.
Additionally, they contain few stickers per conversation, providing limited insight into user preferences or sticker usage patterns. Furthermore, some datasets are not open-sourced, limiting reproducibility and progress of personalized sticker retrieval and recommendation.

\begin{table*}[ht]
\centering
\caption{Comparison with Stickers Datasets. We bold the largest count among publicly available datasets and underline the second largest. Our dataset is cross-domain, large in scale, and contains time and user ID for personalization.}
\begin{tabular}{lccccrc}
\toprule
\textbf{Dataset} & 
\textbf{Pub. Avail.}  & 
\textbf{Cross Domain} & 
\textbf{Time Info.} &
\textbf{\# Sticks./user} & 
\textbf{\# Stickers} & 
\textbf{\# Languages} \\ \hline
StickerTag \cite{stickertag} & \xmark & \xmark & \xmark & \xmark & 13,571 & 1 \\
StickerInt \cite{stickerint} & \xmark & \xmark & \xmark & Unknown & 1,025 & 1 \\
StickerCLIP \cite{stickerclip} & \xmark & \xmark & \xmark & \xmark & 820,000 & 1 \\
CSMSA \cite{CSMSA} & \xmark & \xmark & \xmark & \xmark & 16,000 & 1 \\
SRS, PESRS \cite{learning-to-respond-2021, learning-to-respond-with-stickers-2020} & \xmark & \xmark & \xmark & 6.82 & 320,168 & 1 \\
PerSRV \cite{chee2024persrv} & \xmark & \xmark & \xmark & Unknown & 543,098 & 1 \\ \hline
MCDSCS \cite{mcdscs} & \cmark & \xmark & \xmark & \underline{2.03} & 14,400 & \underline{2} \\
SER30K \cite{SER30K} & \cmark & \xmark & \xmark & \xmark & \underline{30,739} & 1 \\
MOD \cite{MOD} & \cmark & \xmark & \xmark & 1.36 & 307 & \underline{2} \\ \midrule
Ours & \cmark & \cmark & \cmark & \textbf{16.90} & \textbf{370,222} & \textbf{18} \\
\hline
\end{tabular}
\label{tab:dataset_comparison}
\end{table*}

To address these limitations, we introduce U-Sticker, the first large-scale dataset that includes both user information and sticker-based conversations. Our contributions are summarized as follows:
\begin{itemize}
    \item We present U-Sticker, the largest sticker dataset to date, containing \textbf{22K users}, \textbf{370K} stickers and \textbf{8.3M} conversation messages of texts and stickers.
    \item
    U-Sticker is a \textbf{multi-domain dataset} that includes rich and diverse information. Covering 10 domains, it captures temporal, multilingual, and cross-domain behaviors that are not present in previous datasets.
    \item Extensive quantitative and qualitative experiments demonstrate U-Sticker’s practical applications on user behavior analysis and modeling, personalized sticker recommendation. It also holds potential for further research in areas such as personalized retrieval and conversational studies.
\end{itemize}


%% file: Sections/S2-related_work.tex
\section{Related Work}
\label{sec:related_work}

\subsection{Sticker Datasets}
Due to the challenges in preparing sticker conversation-based datasets, their availability remains limited, particularly for publicly accessible datasets. However, recent research on stickers has gained more attention. Currently, there are only nine related datasets, with only \textbf{three} being publicly available. A summary of these datasets is provided in Table \ref{tab:dataset_comparison}.

\subsubsection{Availability}
Many of these datasets are closed-source. Specifically, six datasets are not publicly accessible \cite{stickertag, stickerint, stickerclip, CSMSA, learning-to-respond-2021, chee2024persrv}, meaning they are either closed-source or require additional access that is often unattainable. This lack of open access presents a significant challenge for sticker-based research.

\subsubsection{Cross Domain Information}
Furthermore, none of the aforementioned datasets contain cross-domain information, as the conversations or dialogues are not identified with specific topics. This limitation hinders the exploration of cross-domain sticker usage, which could provide valuable insights.

\subsubsection{User Information}
Additionally, publicly available datasets are scarce in terms of user data, with 1.36 \cite{MOD} and 2.03 \cite{mcdscs} historical stickers per user. This lack of data makes it difficult to perform user analysis and personalization-related tasks. U-Sticker bridges this gap, providing the highest sticker history per user, with an impressive \textbf{16.90} stickers.

\subsubsection{Multilingualism}
Most existing datasets focus on one or two languages, limiting the ability of smaller communities to benefit from advancements in sticker personalization. In contrast, U-Sticker is multilingual, allowing a broader range of communities to engage with these advancements.

As see in Table \ref{tab:dataset_comparison} and above, U-Sticker is the largest available sticker-user dataset. It features \textbf{370,222} stickers, \textbf{cross-domain behavior}, \textbf{rich contextual information}, and \textbf{comprehensive user and temporal} data.




\subsection{Sticker Retrieval}
Most previous research emphasizes the importance of data for sticker retrieval. SRS \cite{learning-to-respond-with-stickers-2020}, PESRS \cite{learning-to-respond-2021} require corresponding utterances, while Lao et al. \cite{practical-sticker} rely on manually labeled emotions, sentiments, and reply keywords. CKES \cite{chen2024deconfounded}annotates each sticker with a corresponding emotion. During sticker creation, Hike Messager \cite{laddha2020understanding} tags conversational phrases to stickers. The reliance on data presents a significant limitation, as stickers without associated information are excluded from consideration.

Gao et al. \cite{learning-to-respond-with-stickers-2020} use a convolutional sticker encoder and self-attention dialog encoder for sticker-utterance representations, followed by a deep interaction network and fusion network to capture dependencies and output the final matching score. The method selects the ground truth sticker from a pool of sticker candidates and its successor. Zhang et al. \cite{zhang-etal-2022-selecting} perform this on recommendation tasks. CKES \cite{chen2024deconfounded} introduces a causal graph to explicitly identify and mitigate spurious correlations during training. The PBR \cite{xia2024perceive} paradigm enhances emotion comprehension through knowledge distillation, contrastive learning, and improved hard negative sampling to generate diverse and discriminative sticker representations for better response matching in dialogues. PEGS \cite{zhang2024stickerconv}, StickerInt \cite{stickerint} generate sticker information using multimodal models and selects sticker responses, but does not consider personalization. StickerCLIP \cite{stickerclip} fine-tunes pretrained image encoders but does not consider personalization. In addition, many methods designed to rank stickers from top-k candidates face a significant drawback in real-world sticker retrieval scenarios, as they quickly become impractical when applied to larger datasets. Possibly due to the lack of dataset, most methods do not consider personalization.

\subsection{Personalized Sticker Recommendation}
The absence of personalized sticker recommendation can likely be attributed to the scarcity of sticker-user datasets. PESRS \cite{learning-to-respond-2021} improves upon earlier work \cite{learning-to-respond-with-stickers-2020} by incorporating user preferences, but the dataset it relies on is not publicly accessible. This limited availability indirectly hampers the development of personalized sticker recommendations and may impede further advancements in the field.





%% file: Sections/S3-method.tex
\section{U-Sticker Dataset}
Telegram is a widely used open-source messaging platform, offering publicly accessible conversations. The interactions within these groups are natural, making them a valuable source for understanding human behavior, particularly in the context of personalization. As such, they present a promising dataset for sticker-related research.

However, the raw data requires preprocessing before it can be utilized effectively. In the following sections, we outline the process of constructing the dataset. We begin by defining the criteria used in the construction process.



\subsection{Construction Criteria}
Given the significant implications of this dataset for future research, we adhere to the following criteria to ensure the quality and reliability of the data source 

We firstly define the envision for the final user-sticker dataset. Since stickers are the focal point of this dataset, we would like a \textbf{high presence of stickers}. Additionally, since we would like to analyze interactions across different domains, the final dataset should involve \textbf{diverse topics}. We would also like the dataset to be of high-quality; avoidance of spam, advertisement and groups with single person. Moreover, we noticed that previous studies overlooked linguistic variety, hence, we wish our final dataset to have \textbf{multilingual coverage}. Finally, to ensure a robust and generalizable dataset, we strive for \textbf{sufficient dataset size}. 

We translate the above goals into the following dataset construction criteria,
\begin{enumerate}
    \item \textbf{Sticker prevalence}: Conversations must contain a significant number of stickers.
    \item \textbf{Topic diversity}: A wide range of discussion topics should be represented.
    \item \textbf{Authentic interactions}: We focus on real user-to-user conversations while avoiding:
    \begin{enumerate}
        \item Announcement-based channels with a single speaker.
        \item Conversations dominated by spam, advertisements, or inappropriate content.
    \end{enumerate}
    \item \textbf{Linguistic diversity}: Conversations should represent multiple languages.
    \item \textbf{Scalability}: We aim to capture as many suitable conversations as possible.
\end{enumerate}

Following these criteria, we manually screened hundreds of conversation groups and ultimately selected \textbf{70} conversation groups to crawl their content.

\subsection{Dataset Pre-processing}
We crawl the data using Telethon \cite{telethon} and process the data in the following sections. Note that users can participate in multiple conversation groups.


\subsubsection{Text Processing}
As our dataset has been collected with multilingual diversity in mind, we identify the languages utilized for ease of downstream processing. We utilize the \textbf{xlm-roberta-base-language-detection} model \cite{conneau2019unsupervised} for language identification. The model provides both the detected language and a confidence score, and we utilize a threshold of 0.99 used to ensure high reliability in language detection.

After language detection, we manually review the classification results, removing entries with fewer than 20 occurrences and those from administrative bots. In total, we detect 18 languages, with the top ten languages in order being English, Russian, French, Spanish, Chinese, Polish, Italian and German, Turkish, Portugese. As can be seen, U-Sticker features 18 languages which is potentially viable for multilingualism sticker-related task.

\subsubsection{Sticker Processing}
The downloaded stickers come in three formats; \texttt{.webp}, \texttt{.webm}, and \texttt{.tgs}, the distribution is 65.6\%, 23.0\% and 11.4\% respectively. We convert all \texttt{tgs} files to the \texttt{.gif} extension using \texttt{tgsconverter} \cite{tgsconverter}, \texttt{.webp} is converted to \texttt{.png} and no conversion is done to \texttt{.webm}. We break up dymamic images into frames and save a frame every second to prepare for later use.

\subsection{Unsafe Text Detection and Replacement}

We aim to moderate the safety of textual content. We perform this in two phases; (1) unsafe text detection and (2) offensive text replacement.

\subsubsection{Unsafe Text Detection}
We identify two types of unsafe text; offensive language and hyperlinks. 

\paragraph{Offensive Language}
Offensive text are content that could contain racism, vulgarities or other potentially offensive content. Since different languages require specialized tools for detecting offensive content, we focus on the ten most used languages in our dataset: \textbf{English, Russian, French, Spanish, Chinese, Polish, Italian and German, Turkish, Portugese}. For each, we employ well-established moderation models that have been published in peer-reviewed conferences, we describe the models in the following and the threshold values. The threshold values were manually selected through stepwise examination of the model outputs, aiming to balance the inclusion of high-quality text with the exclusion of plausibly offensive content.

\begin{itemize}
    \item \textbf{roberta-base-cold} \cite{chinese-moderate}: Detects offensive language in \textbf{Chinese}. The model returns a binary label, positive values are tagged as offensive language.
    \item \textbf{twitter-roberta-base-offensive} \cite{en-moderation}: Identifies offensive language in \textbf{English}. The model returns a 0 to 1 score for offensive and not-offensive tags. We set the offensive threshold as 0.87.
    \item \textbf{russian-sensitive-topic} \cite{russian-content-moderation}: Flags sensitive or offensive content in \textbf{Russian}. The model returns "sensitive" or "not-sensitive" tags. We set "sensitive" content as offensive.
    \item \textbf{toxic-bert} \cite{multi-content-moderation}: A multilingual model that detects offensive language in \textbf{French, Turkish, Portuguese, Italian, Spanish, English, and Russian}. The model returns a score from 0 to 1. We set offensive threshold to be less than 0.0004.
    \item \textbf{dehatebert-mono-polish} \cite{aluru2020deep}: Detects hate speech in \textbf{Polish} language. The model returns a binary label, positive values are tagged as offensive language.
\end{itemize}

\begin{figure}[htbp]
    \centering
    \includegraphics[width=1.0\linewidth]{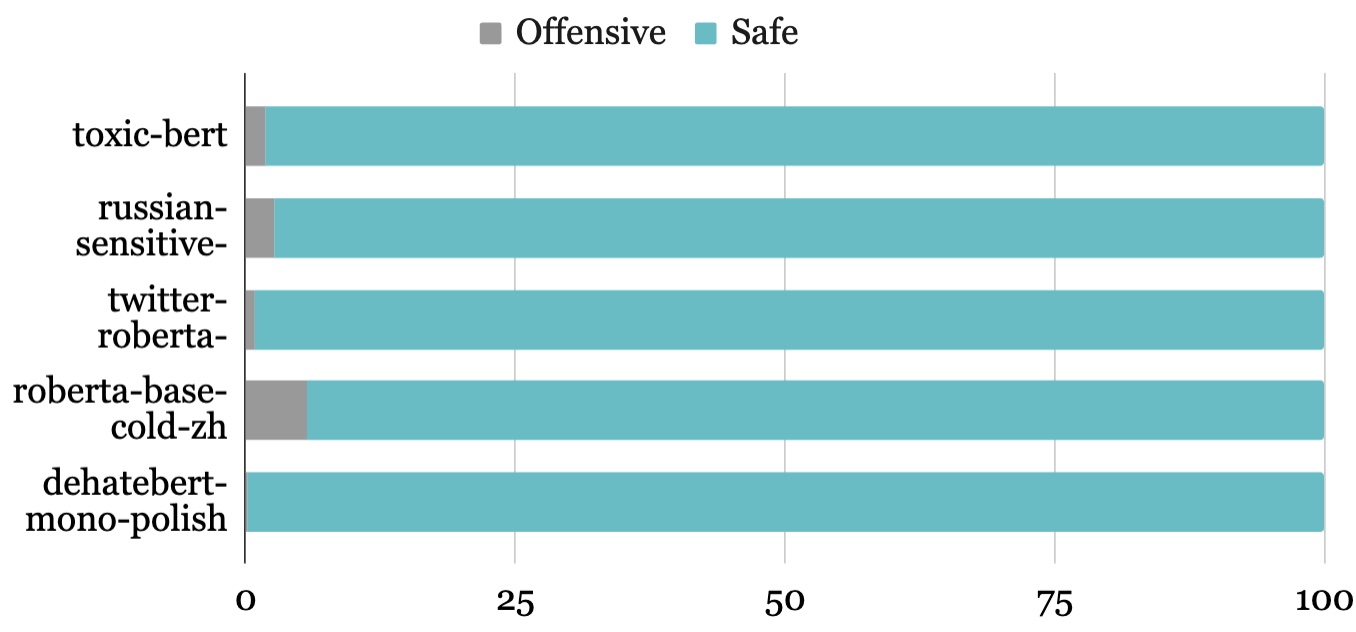}
    \caption{Offensive comment distribution across text classification models. 0.43\% of comments are flagged as offensive.}
    \label{fig:language-detection}
\end{figure}

We present the offensive language detection results in Figure ~\ref{fig:language-detection}.

The dataset shows that offensive language is relatively rare, with only 0.43\% of messages flagged as NSFW. This reflects typical user-generated platforms, where offensive content is present but remains a minority of overall interactions.

\paragraph{Hyperlinks}
Links are naturally present in group conversations. However, these links could potentially contain unsafe content. Hence, we use regular expression \cite{regex} detection to uncover 463,367 links in 313,906 messages and tag them for future modification.

\subsubsection{Unsafe Text Replacement}
\begin{itemize}
    \item \textbf{Offensive Language Replacement}:  
    Offensive or harmful language detected in text messages is replaced with the placeholder label:  
    \textit{U-Sticker detects this as offensive text.}  
    This approach preserves conversation structure while flagging offensive text, ensuring safety without losing context.
    
    \item \textbf{URL/Link Replacement}:  
    When URLs or links are identified in messages, they are replaced with the label:  
    \textit{U-Sticker detects a link.}  
    This ensures private or potentially harmful web addresses are not exposed in the dataset, while preserving the flow of conversation and preventing leakage of sensitive data.
\end{itemize}

\subsection{Unsafe Image Moderation and Replacement}
Next, we move on to unsafe image moderation, where we employ (1) detection and (2) text replacement.

\subsubsection{Unsafe Image Detection}
The purpose of image moderation is to remove offensive images; these include violent, sexual or offensive. To achieve this, we employ several readily available tools. We introduce them in the following;
\begin{itemize}
    \item \textbf{nsfw-image-detection} \cite{falconsai2024nsfw}: has over 54 million downloads on Huggingface and classifies images into normal and not safe for work (nsfw).
    \item \textbf{nsfw-classifier} \cite{giacomo_arienti_2024}: classifies images into four categories - (1) drawings, (2) neutral, (3) hentai and (4) sexy.
    \item \textbf{vit-base-violence-detection} \cite{jaranohaal2024violence}: outputs binary labels for image violence detection.
    \item \textbf{vit-base-nsfw-detector} \cite{adamcodd2024nsfw}: classifies images into nsfw and Safe For Work (sfw).
\end{itemize}


\begin{figure}[htbp]
    \centering
    \includegraphics[width=1.0\linewidth]{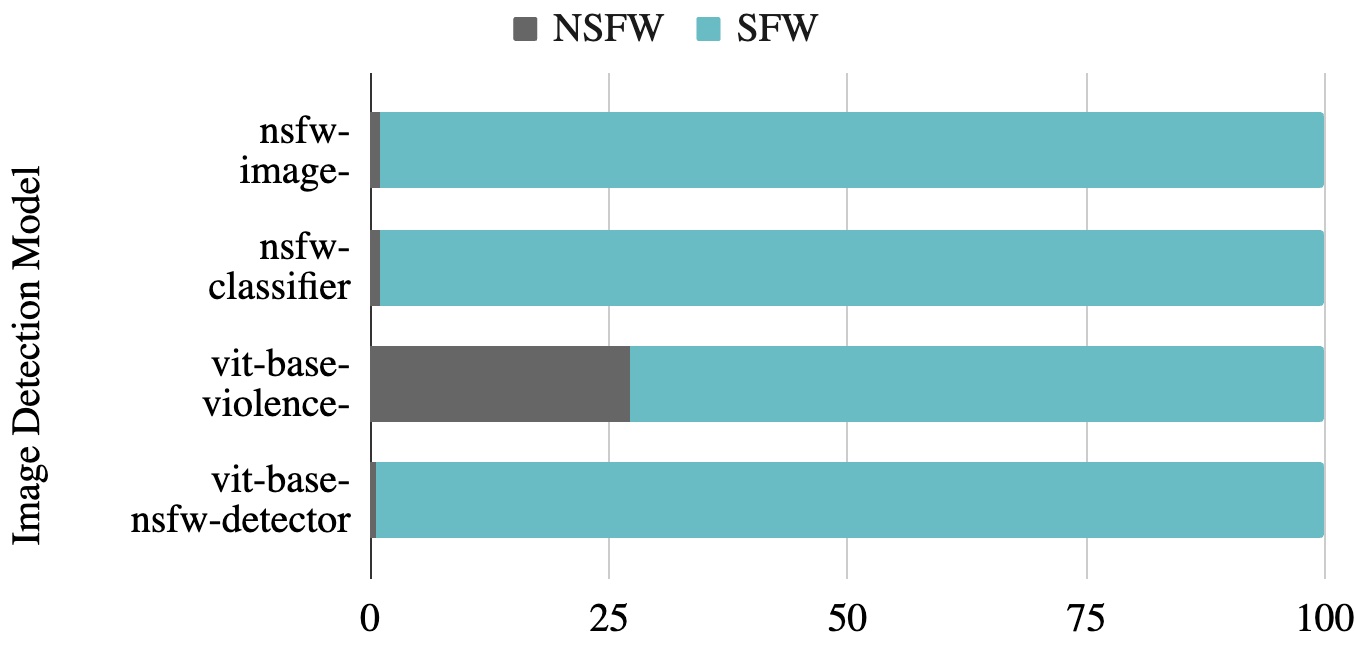}
    \caption{Distribution of NSFW vs. SFW amongst the four offensive image detection models on the raw data. To ensure final labale assignment robustness, we utilize a voting mechanism to balance individual model biases.}
    \label{fig:image-detection}
\end{figure}



We utilize the frame processed above and static images into the four models, where we obtain the nsfw to sfw ratio. We present the results in Figure \ref{fig:image-detection}. nsfw-image-detection is 0.88\%, nsfw-classifier is 0.93\%, vit-base-violence-detection is 27.27\%, vit-base-nsfw-detector  is 0.47\%. 



\subsubsection{Unsafe Image Removal and Replacement}
For the identified offensive images, we replace the textual content with the placeholder \textit{U-Sticker detects this as offensive image}. This helps maintain the integrity of the dataset while ensuring that no harmful visuals, such as explicit or violent content, are retained, thus reducing the risk of privacy violations or inappropriate representations. Then, we remove the sticker entirely from our sticker database.

\subsection{Privatization}
In addition to safety, the privacy of the data is paramount.

\subsubsection{User Identifier}
\paragraph{Hashing sender identifier}
To protect the user's privacy, we anonymize the user identifiers by hashing their original integer identifiers using the SHA-256 hash function \cite{sha256}. Originally, the user identifiers were integers ranging from five-digital value to a maximum value of ten-digit value. After applying the SHA-256 hashing algorithm \cite{sha256}, the user identifiers are transformed into 256-bit strings, which are irreversible and do not retain any direct correlation with the original numeric values. This ensures that the users' identities are kept private, while still allowing for secure and effective data processing and analysis.

\paragraph{Replacing Mentions}
It is common for users to interact with each other in the conversation group. To protect user privacy, we use regular expressions \cite{regex} to identify user identifiers and mentions (e.g., \@username) within text and replace them with the label \textit{\#USER}. Additionally, when users are mentioned via hyperlinks, these occurrences are replaced with \textit{\#USER\_ID}. Importantly, we do not replace the entire utterance but rather focus on replacing only the sensitive information, thus preserving the majority of the conversational context.

\subsubsection{Message Information}
We also anonymize message identifiers through the same SHA-256 hashing algorithm mentioned above.

\subsubsection{Other Sensitive Information}
To prevent the leakage of sensitive user information such as names, ages, addresses, organizations, or phone numbers, we use predefined context dictionaries to search the conversation logs. These logs are then further tagged by Llama-1.5b \cite{liu2023llava}. Any sensitive information found is replaced with the label \textit{\#SENSITIVE-INFORMATION} within the conversation, but are not replaced entirely.


    
    

\subsection{Manual Verification}
After implementing the automatic process unsafe text, image and privatization we conduct a \textbf{manual review} as an additional safeguard to ensure the dataset’s safety and quality. In this phase, a subset of the dataset is randomly sampled and carefully examined by human reviewers. The goal is to verify that the automated procedures have successfully removed or replaced all sensitive and harmful content without affecting the overall structure or meaning of the messages. The manual review process is essential for catching any potential issues that automated processes may have missed, such as false negatives or inaccuracies in labeling. 

Finally, \textbf{370,222 (72.28\%)} of \textbf{512,192} stickers do not contain offensive taggings, \textbf{8,286,422 (99.45\%)} of \textbf{8,332,351} utterances are safe and private. Then, there are a remaining \textbf{67} conversation groups.




\subsection{Conversation Domain Labeling}
To enable groups for future scenario analysis, we categorize conversations into domains. We utilize a two-step approach. First, we read the conversation name, which often provides a clear indication of the conversation's general focus. Next, we browse through the conversation content, examining the topics and discussions that occur within. This process enables us to accurately identify the predominant subject matter of the conversation.

Our categorization is based on domains related to common hobbies and interests. We identify ten primary domains: 

\begin{itemize}
    \item \textbf{Language:} This domain includes conversations focused on learning new languages, sharing language learning resources, or discussing language-related topics.
    \item \textbf{Arts:} Conversations under this domain typically involve the sharing of artwork, discussions around different forms of art, drawing techniques, and other creative activities.
    \item \textbf{Games:} Conversations in this domain revolve around various forms of gaming, whether they are video games, board games, or other interactive entertainment.
    \item \textbf{Technology:} Conversations here engage in code sharing, learning new programming languages, discussing software development, or delving into the latest technological innovations.
    \item \textbf{Finance:} This domain focuses on topics related to financial discussions, including traditional investing, cryptocurrency, blockchain technologies, and financial exchanges.
    \item \textbf{Social:} Conversations under the Social domain are centered around meeting new people, making friends, or organizing social events such as meetups and hangouts.
    \item \textbf{Media Sharing:} Conversations in this category often focus on sharing and discussing media content such as stickers, memes, gifs, and videos for entertainment and social interaction.
    \item \textbf{Outdoor:} Conversations here revolve around nature, outdoor activities like hiking and camping, and the appreciation of animals and the environment.
    \item \textbf{Anime:} This domain is dedicated to discussions about anime, cartoons, and related pop culture topics, including anime recommendations and fan theories.
    \item \textbf{Fan Club:} Conversations in this category are focused on specific celebrities, idols, or fictional characters, where users come together to share their admiration and discuss fan-related topics.
\end{itemize}

While these categories cover a wide range of interests, some topics naturally overlap. For instance, the increasingly popular genre of crypto-based gaming involves elements of both gaming and finance, as users play games to earn cryptocurrency. However, to maintain a concise and organized dataset, we prioritize the dominant domain in such cases—crypto gaming would thus fall under the Finance category in this example. The final dataset, after domain labeling, is presented and analyzed in the following section, where we examine the distribution of conversations across these domains and the insights derived from this classification.

\section{Dataset Analysis}
We analyze the U-Sticker dataset and share our findings in the following section. Firstly, we present a statistical overview of the dataset. Secondly, we dive into the multi-domain characteristics of U-Sticker. Finally, we analyze potential user behavior found in U-Sticker.

\subsection{Statistic}

\subsubsection{Overall Information}

\begin{table}[h]
    \centering
    \caption{Statistics and Comparison of Raw Data and U-Sticker (A) and Top-8 Languages Post-Processing (B).}
    \begin{minipage}{0.68\linewidth}
        \centering
        \caption*{A. Statistics Comparison.}
        \label{tab:dataset-statistics}
        \small
        \begin{tabular}{@{}lcc@{}}
            \toprule
            Field & Raw Data & U-Sticker \\
            \midrule
            \# Domains & - & 10 \\
            \# Groups & 70 & 67 \\
            \# Unique Stickers & 142,981 & 105,803 \\
            \midrule
            \textit{Messages} && \\
            \# Stickers & 512,192 & 370,222 \\
            \# Text & 8,332,351 & 8,286,422 \\
            \midrule
            Avg. Stickers/User & 20.54 & 16.90 \\
            \bottomrule
        \end{tabular}
    \end{minipage}
    \hspace{1pt}
    \begin{minipage}{0.28\linewidth}
        \centering
        \caption*{B. Languages.}
        \label{tab:second-table}
        \small
        \begin{tabular}{@{}c@{}}
            \toprule
            Top-8 Languages \\
            \midrule
            1. English (62.6\%) \\
            2. Russian (9.6\%) \\
            3. French (8.6\%) \\
            4. Chinese (8.0\%) \\
            5. Spanish (2.5\%) \\
            6. Turkish (2.0\%) \\
            7. Vietnamese (1.9\%) \\
            8. Polish (1.5\%) \\
            \bottomrule
        \end{tabular}
    \end{minipage}
\end{table}

    


As shown in Table~\ref{tab:dataset-statistics}, our U-Sticker is large in size and contains user information for cross conversation and cross domain analysis and modeling.
Notably, we filter the raw data from the public platform for safety and privacy concerns to form the final U-Sticker dataset. 

\begin{table}[h]
    \centering
    \caption{Domain Statistics. Notably, Anime, Media Sharing, and Game are the most active domains in terms of sticker usage, while Technology is the least active. The sparsity values highlight the varying levels of engagement across domain. Sparsity values are scaled by a factor of 10,000 for easier readability.}
    \begin{tabular}{@{}lcccccccc@{}} 
    \toprule
    \textbf{Category} &  \textbf{Sticker}& & \textbf{Users}&& \textbf{Sparsity} \\
     & \textit{Global}& \textit{Unique} & \textit{Global} & \textit{Sticker} & \\ \midrule

    Anime (4) & 87,833 & 22,572  & 10,448 & 1,956 & 19.89 \\
    Arts (1) & 219 & 176 & 3.37  & 65 & 191.43 \\
    Fan Club (4) & 2,287 & 1,234  & 6,341 & 168 & 110.32 \\
    Finance (8) & 45,655 & 10,083 & 245,020 & 7,894 & 5.74 \\
    Game (1) & 79,367 & 16,382 & 48,297 & 2,468 & 19.63 \\
    Language (5) & 51,556 & 15,801 & 50,734 & 3,456 & 9.44 \\
    Outdoor (5) & 2,088 & 953 & 6,230 & 265 & 82.68 \\
    Social (9) & 13,077 & 3,894 & 42,180 & 1,176 & 28.56 \\
    Media (18) & 87,827 & 45,451 & 14,309 & 7,498 & 2.58 \\
    Tech (2) & 313 & 210 & 4,425 & 75 & 198.73 \\
    \bottomrule     
    \end{tabular}
    \label{tab:category-language-sparsity}
\end{table}

\subsubsection{Sticker Usage Distribution}
\begin{figure}[htbp] 
    \centering \includegraphics[width=1.0\linewidth]{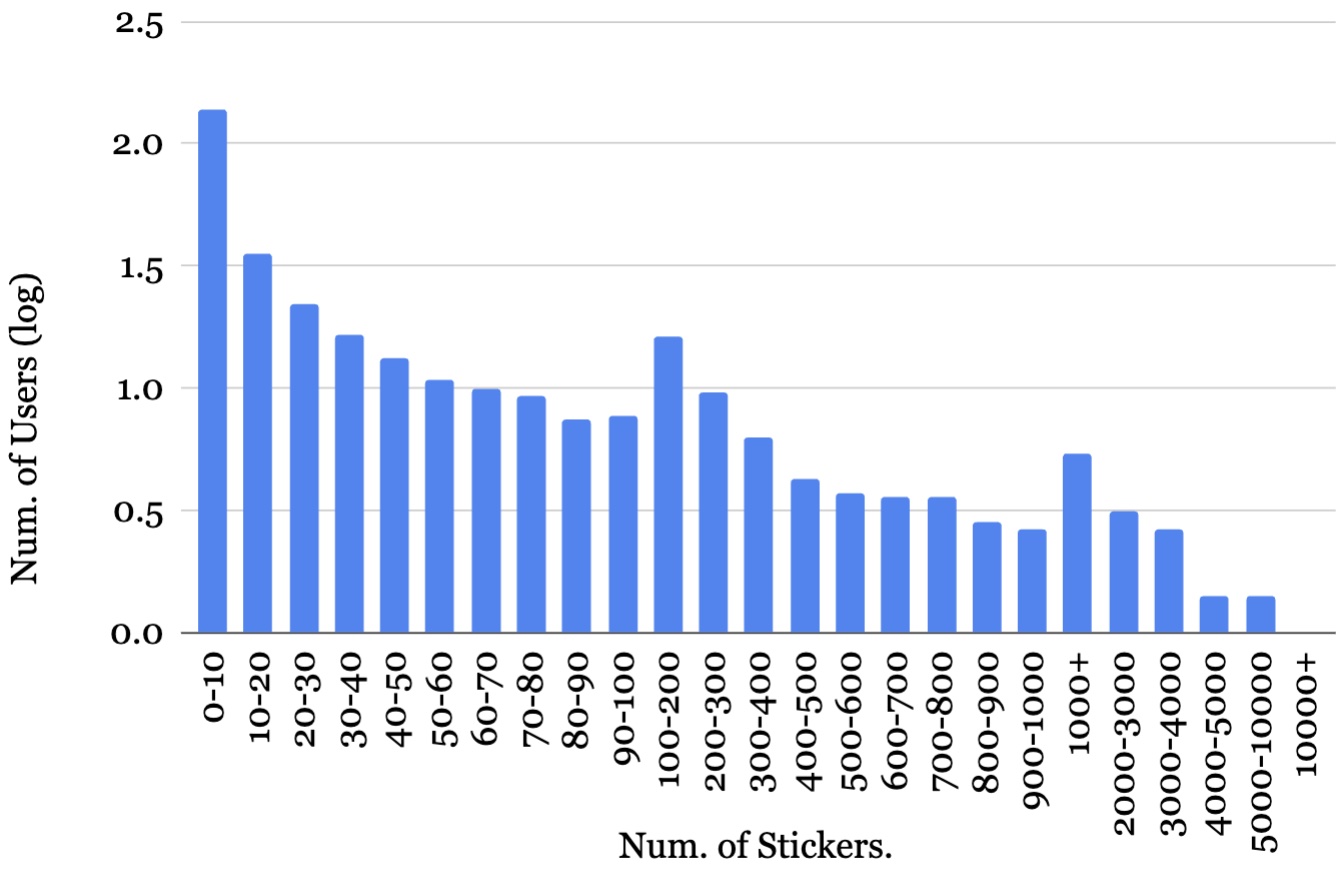} 
    \caption{Sticker Usage Distribution. Number of users against the number of stickers. U-Sticker usage trend follows a long-tail distribution, highlighting the real-world sticker usage patterns.}
\label{fig:stickers-user-distribution} 
\end{figure}

Next, we analyze the distribution of sticker usages. To better visualize this distribution and prevent distortion, we apply a logarithmic transformation to the number of users. As demonstrated in Figure ~\ref{fig:stickers-user-distribution}, the U-Sticker usage trend follows a long-tail distribution, where most users employ between 0 and 100 stickers, while only a small fraction exhibit extremely high usage. This behavior is consistent with Zipfian or Pareto-like distributions \cite{zipf1949human} \cite{newman2005power}, and is typical of organic user activity observed in large-scale, real-world datasets.s.

\subsubsection{Language Distribution}
Then, we analyze the language distribution of U-Sticker, shown in Figure ~\ref{tab:second-table}. U-Sticker features Arabic, Bulgarian, Chinese, Dutch, English, French, German, Hindi, Italian, Japanese, Modern Greek, Polish, Portuguese, Russian, Spanish, Swahili, Thai, Turkish, Urdu, and Vietnamese. English is the most dominant language with an domineering 62.6\%. Russian (9.6\%) is the second most popular language followed by French (8.6\%) and Chinese (8.0\%). 


\subsection{Multi-Domain Characteristics}
In this section, we analyze the multi-domain characteristics of U-Sticker. Firstly, we analyze the domain-group distribution. Secondly, we introduce the domain statistics. Lastly, we analyze the user-sticker usage overlap across domains.


\subsubsection{Group Distribution}
As mentioned above, U-Sticker contains 67 groups and 10 domains, with the median group being five. The minimum group count is one, being Arts, and the maximum group count is media sharing. This is due to the nature of resource, focusing on stickers. Games and Social come close as they tend to be relaxed and so more expressed.


\subsubsection{Domain Statistics}
We present the overview statistics of domain in Table \ref{tab:category-language-sparsity}. Looking at sticker global counts, the domain with the most sticker is Anime, followed by Media Sharing and Game. This is reasonable as these are highly relaxed groups and tend to be more expressive, thereby having a higher sticker usage count.

The most sparse domain is Technology and the most dense domain is Media Sharing. This is natural as technology is a rather technical topic and audience could be less inclined to using sticker expressions. On the other hand, media sharing sometimes encourage chats entirely comprising of stickers, hence increasing the density. Surprisingly, Finance and Language are also denser domains. This could be because Finance and Language are semi-professional scenarios which could call for an increased sticker usage.

Lastly, the average stickers used per user varies largely with Anime and then game domain having the highest 44.90 and 32.16 respectively. On the other hand, Technology and Finance domains have 4.17 and 5.78 stickers respectively. We hypothesize that the major difference in the nature of the scenario affects sticker usage in users.


\subsubsection{User Sticker Usage Overlap Across Domains}
\begin{figure}[htbp] 
    \centering \includegraphics[width=1.0\linewidth]{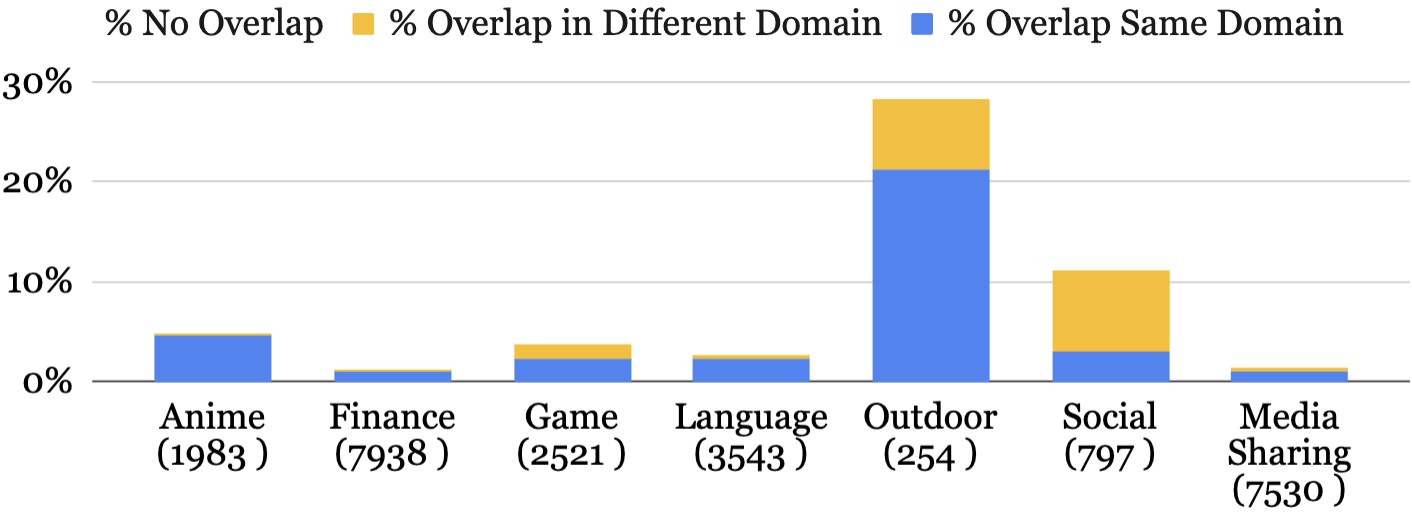} 
    \caption{Distribution of users' sticker usage. Social domain users features the largest distribution of users in contrastive (yellow), reflecting the inherently explorative nature of their domain.}

\label{fig:sticker-overlap-distribution} 
\end{figure}

Lastly, we analyze the appearance of users in the same domain but in other groups, or in entirely different domain. We plot our findings in Figure \ref{fig:sticker-overlap-distribution}.

Firstly, we establish that users do join other groups and participate; these could be within the same domain or an entirely different domain. The number beside the label on the horizontal axis reveals the total number of sticker users within the domain, the blue bar represents the percentage of users that appear in other groups within the same domain. We omit domains with fewer than 50 user stickers to avoid generalization.

Secondly, two of the largest groups with more than 7000 sticker users (Finance and Media Sharing domain) with roughly 15\% and 3\% to 85\% and 97\% different to same domain ratios. In fact, other than the Social domain, domains all have more users participating in the same domain than somewhere else; which could possibly reveal user preference and taste. Naturally, Social domain is clearly an explorative domain and therefore behaves as such.

\section{Application Case 1: User Behavior Analysis}

Next, we move on to the user behavior analysis on the U-Sticker dataset. More specifically, we demonstrate U-Sticker's potential on showcasing (1) user distinct style (2) multi-domain behavior.

\subsection{Distinct User Styles}

\begin{figure}[htbp] 
    \centering \includegraphics[width=0.7\linewidth]{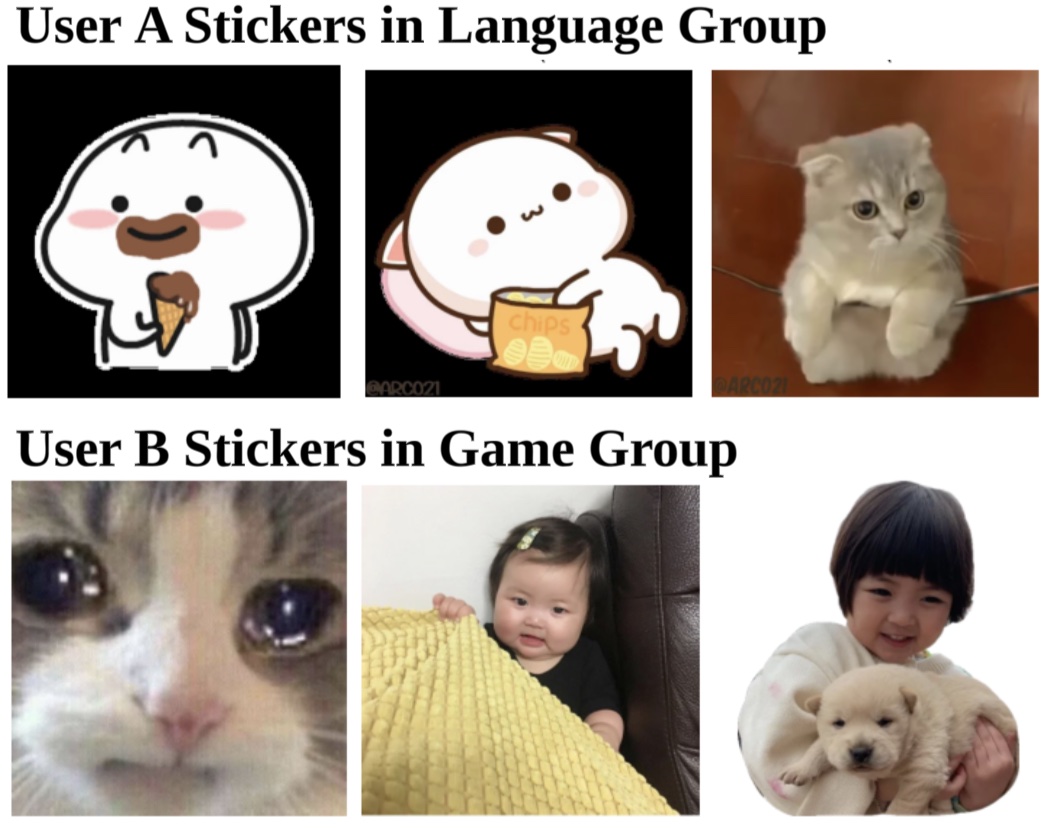} 
    \caption{User A stickers in the Language group and User B stickers in the Game group.} 
\label{fig:sticker-user-profile-B} 
\end{figure}

As shown in Figure \ref{fig:sticker-user-profile-B}, User A's sticker styles are predominantly cat-themed and cute, while User B’s stickers are more idol-based, with an emphasis on adorable and charming designs. This demonstrates that users generally have a distinct sticker preference and this potentially be used for user modeling.


\subsection{Cross Domain Behavior Change}
\begin{figure}[htbp] 
    \centering \includegraphics[width=0.7\linewidth]{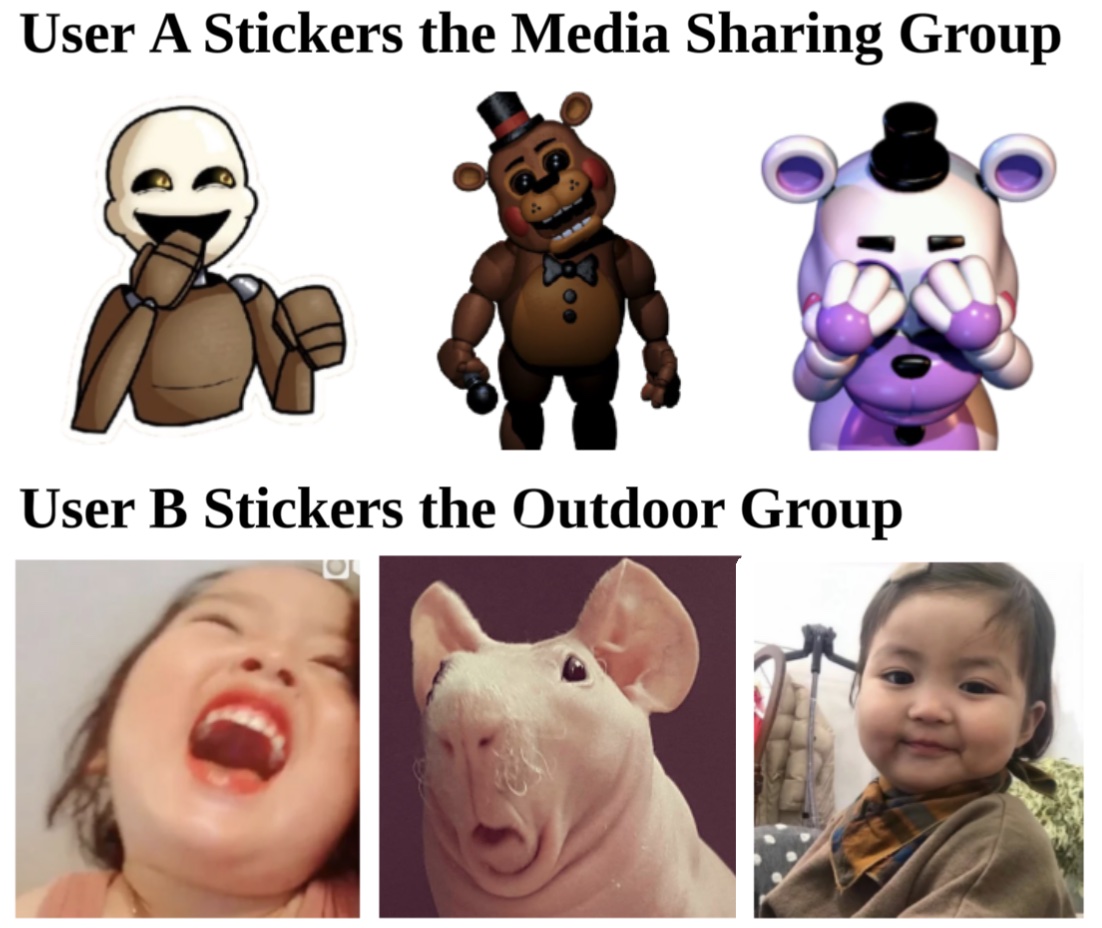} 
    \caption{User A stickers in the Media Sharing group and User B stickers in the Outdoor group.} 
\label{fig:sticker-user-profile-A-2} 
\end{figure}


More interestingly, we analyze if user, due to different context, could change their style. Herein, the same User A, Figure \ref{fig:sticker-user-profile-A-2}, displays a significantly different style in the Media Sharing group, which is more structured and cartoonish in appearance, vastly contrastive to the originally cartoonish style. On the other hand, User B’s stickers, still retain elements of their signature adorable style but are subtly tuned to match the outdoor-themed context of the new domain, which is about rats. This behavior reveals that different user have different consistency level in sticker-usage. There are possibly many factors for this behavior such as the group context. 

\subsection{Temporal Behavior Change}



\begin{figure}[htbp] 
    \centering \includegraphics[width=0.7\linewidth]{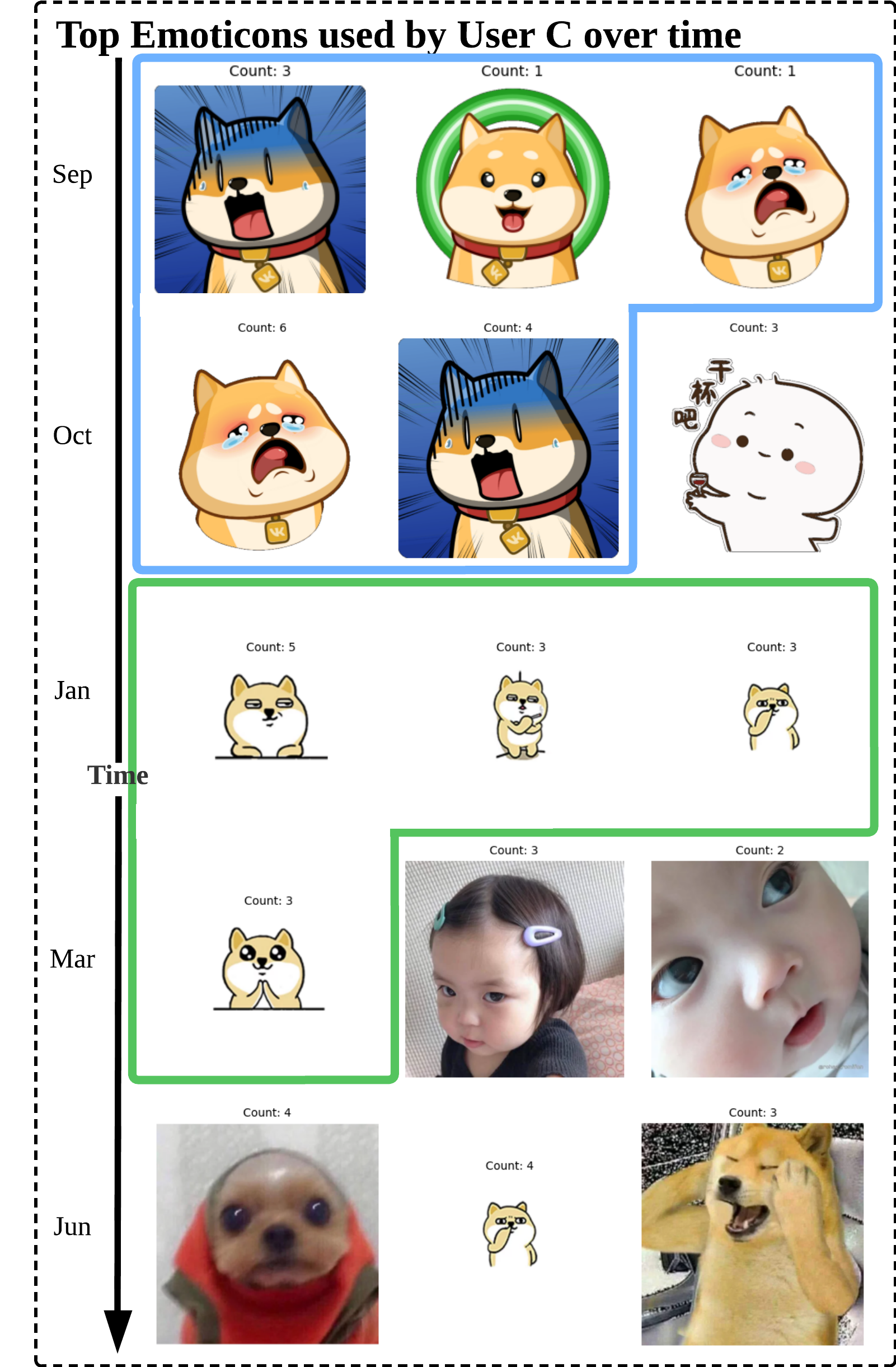} 
    \caption{User C’s evolving sticker preferences over time, shifting from expressive dog stickers in September to smaller, more reserved dog stickers by January, with a renewed interest in realistic dog stickers by June, maintaining a consistent preference for dog-related designs.} 
\label{fig:user-overtime-change} 
\end{figure}

We present the top-used stickers of User C in Figure ~\ref{fig:user-overtime-change}, showing how their preferences shift over time. In September and October, they favor cartoonish dog stickers; by January, they prefer smaller, cuter dog designs; and in June, they lean toward realistic dog stickers while still using the cute ones. Despite style changes, their consistent preference for dog-themed stickers—especially cute ones—remains.

\begin{figure}[htbp] 
    \centering \includegraphics[width=0.5\linewidth]{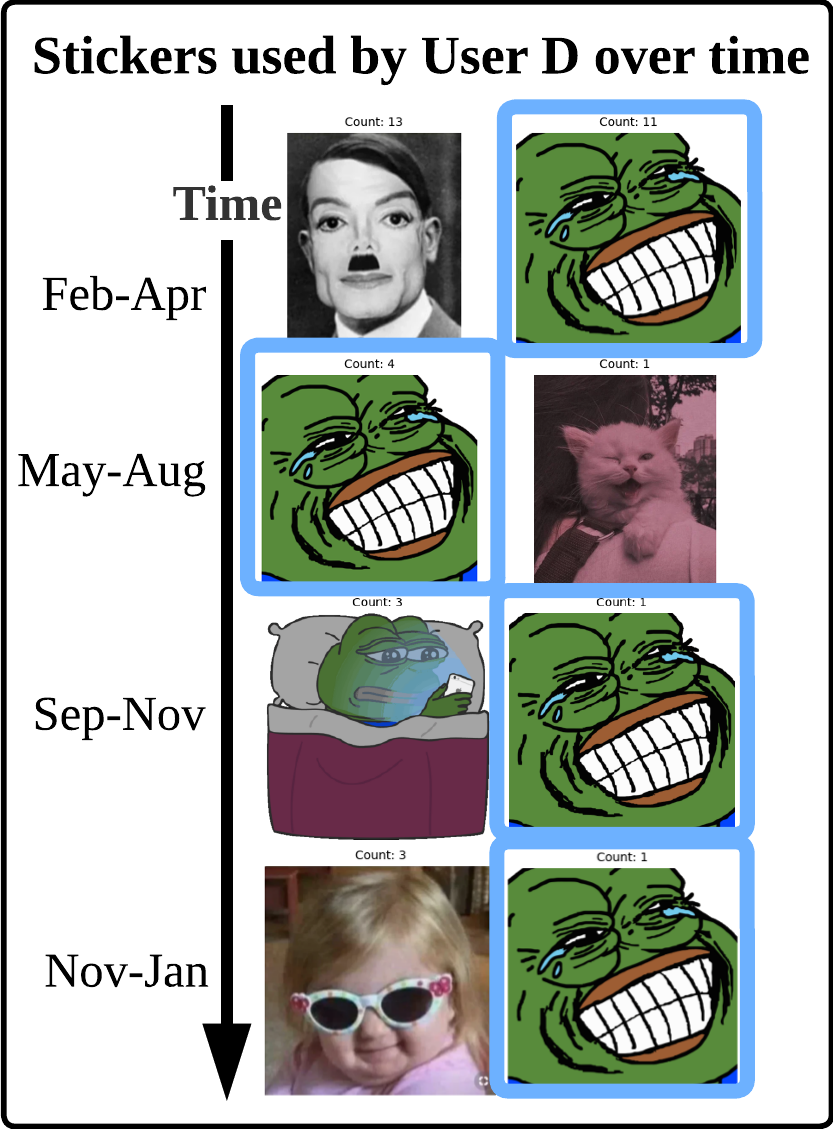} 
    \caption{User D’s consistent usage of the green laughing frog sticker (Pepe frog) from February to January, with occasional months of no sticker usage, highlighting a strong preference for this particular sticker over time.} 
\label{fig:no-change-overtime} 
\end{figure}

In contrast, User D, Figure ~\ref{fig:no-change-overtime} consistently uses the same green laughing frog sticker (commonly known as "Pepe") across many months, with few gaps. Overall, this demonstrates that our dataset contains both types of users: those like User D, who exhibit stable preferences for specific stickers, and those like User C, whose preferences evolve over time. Intuitively, this aligns with common sense, as people can have both consistent and fluctuating preferences depending on various factors. This diversity in our dataset enhances the accuracy and realism of the personalized sticker retrieval task, allowing it to better reflect the range of behaviors found in real-life scenarios.

\subsection{User Behavior with Different Responders}

\begin{figure}[htbp] 
    \centering \includegraphics[width=0.8\linewidth]{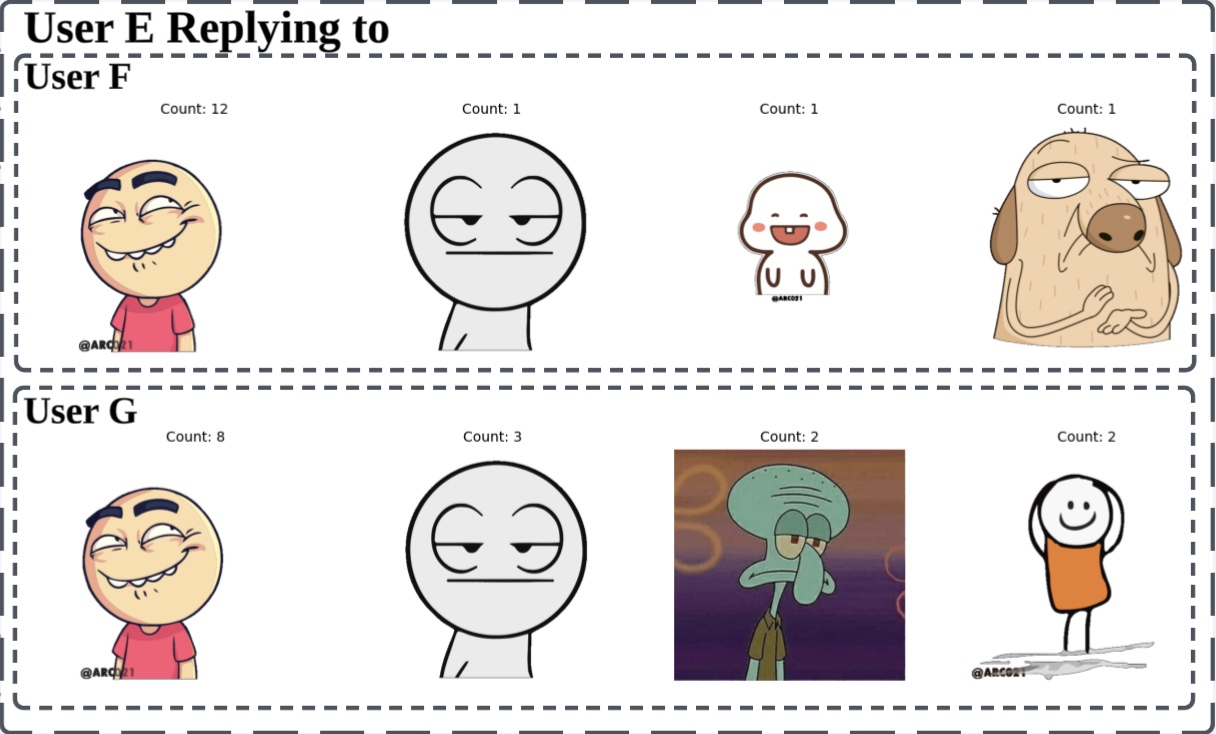} 
    \caption{User E consistently uses the red t-shirt with smirk boy sticker and the grey cartoon straight-face sticker when replying to Users F and G} 
\label{fig:same-sticker-diff-user} 
\end{figure}

We highlight two examples of how sticker usage varies with different recipients. In the first case, Figure~\ref{fig:same-sticker-diff-user}, User E replies to both Users F and G using a consistent set of stickers—primarily a nonchalant boy sticker—demonstrating stable preferences regardless of recipient.

\begin{figure}[htbp] 
    \centering \includegraphics[width=1.0\linewidth]{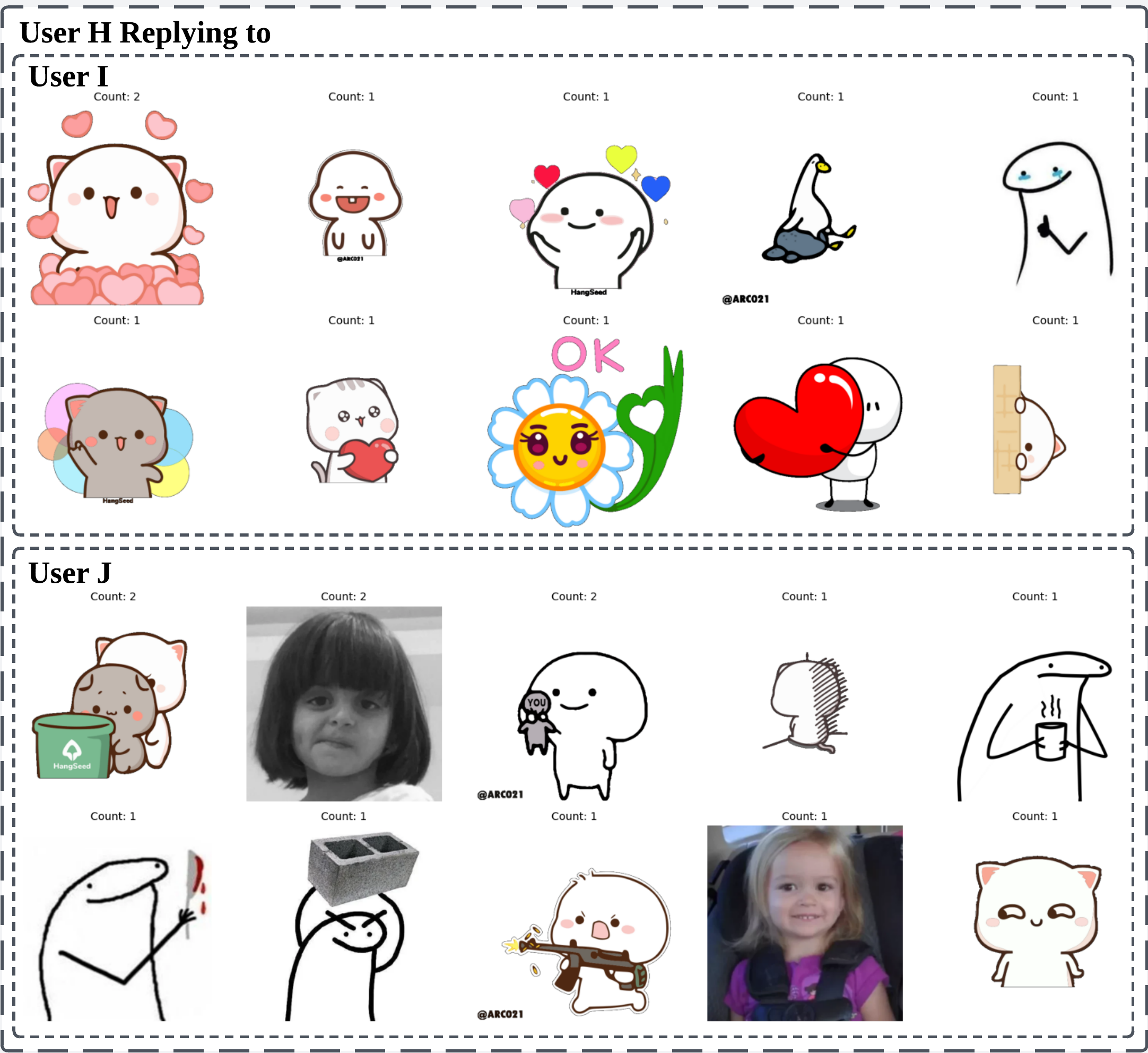} 
    \caption{User H displays contrasting style preferences when replying to Users I and J, using cute, cuddly bear stickers for User I and black-and-white, melancholic stickers for User J.} 
\label{fig:diff-user-sticker-diff} 
\end{figure}

Contrastively, User H adapts their sticker style depending on the recipient: using cheerful, colorful stickers with User I, but switching to melancholic, monochromatic stickers with User J, Figure~\ref{fig:diff-user-sticker-diff}.

In conclusion, these examples demonstrate that our dataset captures meaningful variations in sticker usage based on the context of the interaction. This variability, driven by the recipient, intuitively aligns with our commonsense understanding of how individuals communicate differently with different people. U-Sticker effectively captures this behavior, which in turn empowers related tasks, such as personalized sticker retrieval, by enabling a deeper and more context-aware analysis of user preferences.

%% file: Sections/S4-experiments.tex
\section{Application Case 2: Sticker Recommendation}
\label{sec:experiments}
In this section, we evaluate the feasibility of our dataset through quantitative experiments. Specifically, we evaluate U-Sticker dataset on current sticker recommendation.


\subsection{Experiment Settings} 
We evaluate two baseline methods for sticker recommendation, described as follows:

\begin{enumerate} 
    \item \textbf{MOD} \cite{MOD}: Takes a list of negative samples along with one true positive sample and calculates the probability of each sample being the correct match. The results are then used to rank the samples.
    \item \textbf{SRS} \cite{learning-to-respond-2021}: This approach ranks a list containing negative and one positive sample, returning the ranked results.
\end{enumerate}

We conduct experiments on three distinct perspectives of the U-Sticker dataset: (1) the English and Chinese subsets, (2) the domain-specific subset, and (3) the complete dataset.

Since MOD can only handle English and Chinese, we select groups with more than 80\% of text being either English or Chinese and train the model on it. We do the same for SRS. This forms the \textbf{English and Chinese Subset} setting. To interpretate the performance between different domains, we separate the dataset by domains as shown in Table \ref{tab:category-language-sparsity}. Since SRS is not limited by multilingualism, we evaluate using SRS. This forms the \textbf{Domain} setting. Finally, the \textbf{complete} U-Sticker dataset is evaluated on SRS.

\subsection{Results}
\subsubsection{English and Chinese Subset}
Based on the experimental results presented in Table \ref{tab:experiments-language}, we observe that the English language performs better than the Chinese language across both the MOD and SRS methods. Specifically, the English subset consistently shows higher scores in all evaluation metrics (MRR, R@1, R@3, R@5, and R@10). For instance, in the 1:9 ratio, the MOD method for English achieves an MRR of 0.289, while the Chinese MOD method only reaches 0.317, this could be due to the smaller data size of Chinese in comparison to English.

\begin{table}[htbp]
\centering
\caption{Sticker Retrieval Results for MOD and SRS on the English and Chinese. Negative Samples is shortformed NS.}
\begin{tabular}{lclccccc}
\toprule
Lang & NS & Method & MRR & R@1 & R@3 & R@5 & R@10 \\ \midrule
 \textit{En} & 1:9 & MOD  & 0.289 & 0.307 & 0.495 & 0.224 & 1.000  \\
  & & SRS  & 0.291 & 0.098 & 0.298 & 0.495 & 1.000 \\  \cmidrule(lr){2-8} 
 &  1:19 & MOD & 0.180 & 0.051 & 0.145 & 0.250 & 0.511 \\ 
 &  & SRS & 0.299  & 0.050 & 0.155 & 0.256 & 0.429 \\ \hline
 
 
\textit{Zh}  &  1:9 & MOD  & 0.317 & 0.125 & 0.344 & 0.531 & 1.000  \\
& & SRS   & 0.264 & 0.056 & 0.281 & 0.500 & 1.000 \\ \cmidrule(lr){2-8} 
 &  1:19 & MOD & 0.156 & 0.063 & 0.094 & 0.156 & 0.406  \\ 
 &  & SRS & 0.270 & 0.061 & 0.138 & 0.219 & 0.377  \\


 
\bottomrule
\end{tabular}
\label{tab:experiments-language}
\end{table}

\subsubsection{Domain}
Table \ref{tab:domain-negative-samples} reveal that the \textbf{Media Sharing} domain performs the best overall, with an MRR of 0.335, R@1 of 0.143, R@3 of 0.365, and R@5 of 0.552. On the other hand Arts domain has a relatively lower performance, possibly due to the higher sparsity (0.019143).

\begin{table}[htbp]
    \centering
    \caption{Results of SRS on individual domains. Media Sharing domain performs the best overall, whereas Arts domain has a relatively lower performance possibly due to higher sparsity.}
    \begin{tabular}{lllll}
    \toprule
    \textbf{Domain} & \textbf{MRR} & \textbf{R@1} & \textbf{R@3} & \textbf{R@5} \\
    \hline
    
     \textbf{Anime} 
     & 0.294 & 0.101 & 0.298 & 0.500 \\
    
    \textbf{Arts} 
   & 0.288 & 0.094 & 0.281 & 0.484 \\

    \textbf{Fan Club}
     & 0.280 & 0.080 & 0.280 & 0.510 \\

    \textbf{Finance} 
    & 0.296 & 0.080 & 0.280 & 0.510 \\
    
    \textbf{Game} 
     & 0.293 & 0.100 & 0.301 & 0.498  \\

     \textbf{Language} 
     & 0.296 & 0.103 & 0.307 & 0.501 \\
     
    \textbf{Outdoor} 
    & 0.298 & 0.103 & 0.324 & 0.500 \\
     
     \textbf{Social}
    & 0.300 & 0.105 & 0.305 & 0.513 \\
    
    \textbf{Media Sharing} 
    &  0.335 & 0.143 & 0.365 & 0.552 \\
    \textbf{Tech} 
    & 0.321 & 0.125 & 0.361 & 0.611  \\
   
    \bottomrule
    \end{tabular}
    \label{tab:domain-negative-samples}
\end{table}

\subsubsection{Complete Dataset}
The results in Table \ref{tab:srs-full} show that performance decreases as the number of negative samples increases. With 4 negative samples, the model achieves a \textbf{MRR} of 0.425 and \textbf{R@1} of 0.202, indicating good performance. As the number of negative sample increase, a decline in accuracy is observed, which is consistent with general recommendation models. 
\begin{table}[htbp]
\centering
\caption{Performance metrics for varying NS across multiple evaluation criteria on the entire dataset using SRS method.}
\begin{tabular}{ccccccc}
\hline
\multirow{2}{*}{{NS}} & \multicolumn{6}{c}{Evaluation Metrics} \\
\cline{2-7}
& MRR  & R@1  & R@3  & R@5  & R@10 & R@20 \\
\hline
1:4  & 0.425 & 0.202 & 0.603 & 1.0  & -    & - \\
1:9  & 0.294 & 0.102 & 0.303 & 0.499 & 1.0  & - \\
1:19 & 0.296 & 0.054 & 0.153 & 0.251 & 0.332 & 1.0 \\
\hline
\end{tabular}
\label{tab:srs-full}
\end{table}

\section{Discussions}

\subsection{Other Potential Applications of U-Sticker}
Beyond user behavior analysis and sticker recommendation, the U-Sticker dataset presents several opportunities for further research. One potential avenue is a more extensive \textbf{User Behavior Modeling}, which could involve more quantitative user analysis methods, or involve analyzing temporal changes in user behavior and investigating whether sticker usage varies based on the responder. Additional potential applications include, but are not limited to, \textbf{Personalized Sticker Retrieval} and \textbf{Generative AI}, both of which have previously faced dataset limitations \cite{pmg, pigeon}.

We acknowledge that there are several limitations in our work. Despite extensive automatic and manual data verification, privacy and safety concerns may still arise, presenting an opportunity for advancements in addressing such issues.

\subsection{Ethical Considerations}
We affirm that we have used the data in accordance with Telegram's Terms and Conditions \cite{telegram_tos}. Given a user's expression of opposition, we refrain from crawling to respect their privacy. Additionally, our dataset aligns with the FAIR principles, which emphasize making data \textit{Findable}, \textit{Accessible}, \textit{Interoperable}, and \textit{Reusable} \cite{fair_principles}.

\section{Conclusion}


In this paper, we introduce U-Sticker dataset, the largest sticker dataset to date, comprising data from 22.6k users, 370.2k stickers, and 8.8M conversation messages. This multi-domain dataset encompasses a wealth of diverse information, capturing temporal, multilingual, and cross-domain behaviors not found in previous datasets. 
Extensive quantitative and qualitative experiments demonstrate the practical applications of U-Sticker in user behavior modeling and personalized sticker recommendation. It also holds potential for further research in areas such as personalized retrieval and conversational studies.

%% file: Sections/S5-appendix.tex
\onecolumn